\documentclass{ifacconf}

\usepackage{graphicx}      
\usepackage{natbib}        
\usepackage{amsmath} 
\usepackage{amssymb}
\usepackage{tabls}
\usepackage{mathtools}
\usepackage{float}

\usepackage{blkarray, bigstrut}
\usepackage{esvect}
\usepackage{cuted}
\usepackage{algorithmicx}
\usepackage{tikz}
\usepackage{algorithm}
\usepackage{algpseudocode}
\usepackage{multicol,lipsum}
\algrenewcommand\algorithmicrequire{\textbf{Input:}}
\algrenewcommand\algorithmicensure{\textbf{Output:}}
\makeatletter
\def\BState{\State\hskip-\ALG@thistlm}
\DeclareMathOperator*{\argminA}{arg\,min} 

\makeatother
\newcommand\norm[1]{\left\lVert#1\right\rVert}
\graphicspath{{./Figs/}}
\begin{document}
\begin{frontmatter}

\title{Sparse Identification of Nonlinear Duffing Oscillator From Measurement Data} 

\author[First,Second]{S. Khatiry Goharoodi} 
\author[First]{K. Dekemele} 
\author[First]{L. Dupre} 
\author[First]{M. Loccufier} 
\author[First,Second]{G. Crevecoeur} 
\address[First]{Department of Electrical Energy, Metals, Mechanical Constructions and Systems (EEMMeCS), 
Ghent University, Technologiepark 913 and 914, B-9052 Zwijnaarde, Belgium \\
(e-mail: Saeideh.KhatiryGoharoodi@ugent.be)
}
\address[Second]{Flanders Make, the strategic research centre for the manufacturing industry in Flanders, Belgium}

\begin{abstract}                
In this paper we aim to apply an adaptation of the recently developed technique of sparse identification of nonlinear dynamical systems on a Duffing experimental setup with cubic feedback of the output. 
The Duffing oscillator described by nonlinear differential equation which demonstrates chaotic behavior and bifurcations, has received considerable attention in recent years as it arises in many real-world engineering applications. Therefore its identification is of interest for numerous practical problems. 
To adopt the existing identification method to this application, the optimization process which identifies the most important terms of the model has been modified. 
In addition, the impact of changing the amount of regularization parameter on the mean square error of the fit has been studied.
Selection of the true model is done via balancing complexity and accuracy using Pareto front analysis. 
This study provides considerable insight into the employment of sparse identification method on the real-world setups and
the results show that the developed algorithm is capable of finding the true nonlinear model of the considered application including a nonlinear friction term.
\end{abstract}


\end{frontmatter}

\section{Introduction}

The cubic Duffing equation as a differential equation with third-power nonlinear term is an example of a dynamic system that exhibits chaotic behavior and bifurcations. The Duffing equation arises in a vast number of real-world applications in various fields of engineering and physics. 

Applications in secure communication and feature extraction from weak mechanical fault signal have been recently discussed in \cite{liu2015observe} and \cite{han2016feature} respectively. The usage of the Duffing in real-time image encryption has been proposed in \cite{kuiate2017autonomous}. Moreover several papers such as \cite{lai2016weak} used Duffing for weak-signal detection.
In addition many attempts have been made (\cite{das2016energy}, \cite{rice1987practical}) to investigate the nonlinear vibration absorbers which have been designed base on Duffing-type equations.

In literature there have been several attempts to identify the Duffing system in which considerable amount of them are focused on estimation of the parameters. 
In \cite{gandino2010identification} the nonlinear subspace identification method combined with different excitation inputs has been applied on an oscillator describing Duffing equation to identify its parameters. Moreover, the parameters of an experimental beam shaker presenting a Duffing-type system have been identified based on measuring the jump-down frequencies by \cite{tang2016using}, while in \cite{quaranta2010parameters} a Van der Pol-Duffing oscillator's parameters have been evaluated by means of particle swarm optimization algorithms.
Other examples of parameter identification of Duffing system can be found in \cite{kerschen2006past}.

Black-box identification of the Duffing equation has been investigated as well.
The Mathematical models of the damped Duffing have been determined by \cite{masri1993identification} using artificial neural networks. 
A similar approach was followed in \cite{masri2004identification} where a parametric model identification was proposed based on a set of basis functions and applying least-squares.

The recent method of sparse identification of nonlinear dynamics by \cite{brunton2016sparse} follows a similar procedure associated with the compressed sensing theory (\cite{candes2008introduction}). This methodology starts with the construction of a library based on a combination of linear and nonlinear basis functions of the measurements and the input. Subsequently the sequential thresholded least-squares algorithm is applied to find the true model of the system, depending on choosing the accurate value of the regularization parameter. The approach has been successfully applied on several sets of simulation data without assumptions on the form of the governing equations.

In this paper a revised version of sparse identification using the alternating direction method of multipliers (ADMM) is for the first time to the authors' knowledge applied on a set of real-world experimental data. The investigated experimental setup here was specifically designed as a passive vibration absorber. 
In \cite{dekemele2018performance}, the importance of viscous damping, linear and cubic stiffness coefficients in vibration absorbing performance was shown. Therefore, it is vital to correctly identify these terms.

This paper is divided into five sections.
The second section gives a brief overview of the Duffing oscillator setup.
Next, the identification algorithm is described followed by the results and discussion that are presented in Section 4, showing the ability of the technique to find the actual model.
Our conclusions are drawn in the final section.

\section{Duffing oscillator setup}

\begin{figure}[b]
	\centering
	\includegraphics[width=8.5cm]{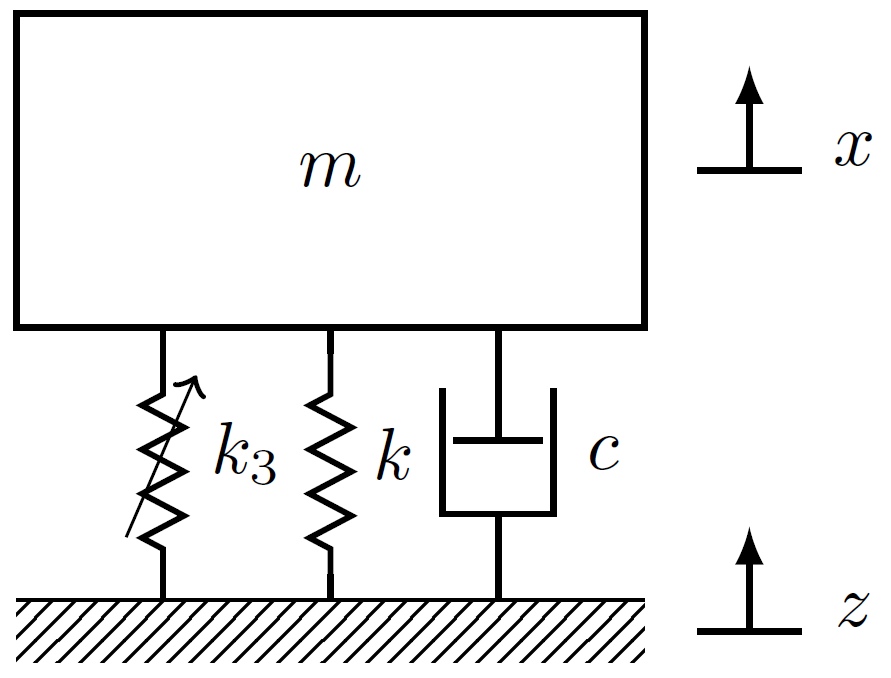}
	\caption{The Duffing oscillator subjected to ground excitation}
	\label{fig:duff}
\end{figure}

The experimental setup imposes a base excitation $z$ to a Duffing oscillator, see Fig. \ref{fig:duff}, with coordinate $x$ as the displacement of the mass.
The dynamics of the base-excited Duffing oscillator are

\begin{equation}
m\ddot{x}= - c\left(\dot{x}-\dot{z}\right) -k\left(x-z\right) - k_{3}\left(x-z\right)^3
\end{equation}

or expressed in the relative ground coordinate $q \triangleq x - z $:

\begin{equation}
m\ddot{q}= - c\dot{q} - kq - k_{3}q^3 -m\ddot{z}
\label{eq:Duffing_ground}
\end{equation}
\subsection{Design principle}

\begin{figure}
	\centering
	\includegraphics[width=8.4cm]{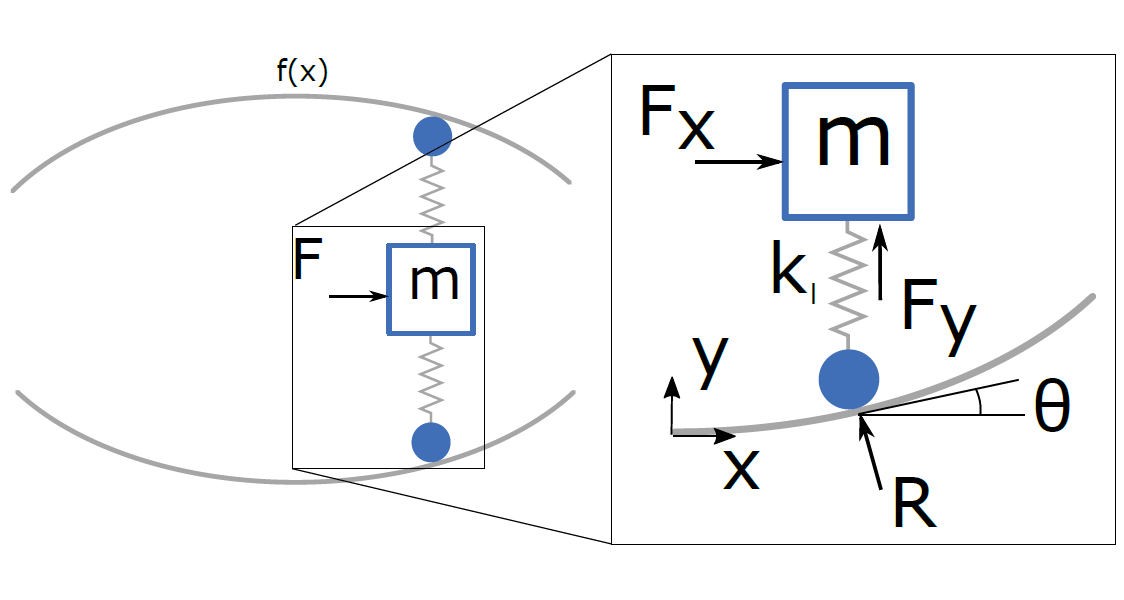}
	\caption{Design principle of the Duffing oscillator}
	\label{fig:principle}
\end{figure}

A horizontal mass-spring system is in equilibrium, see Fig.~\ref{fig:principle}. To both springs with linear stiffness $k_{l}$, a follower is attached. As the mass moves horizontally, the followers move according to a \emph{profile} $f(x)$, which compresses the springs.
Besides the static horizontal force $F_{x}$, which represents the stiffness characteristic in the horizontal direction, there is also a reaction force caused by the spring $F_{y}$ on the mass. The profile imposes a reaction force $R$ on the followers. These 3 forces are related by:

\begin{eqnarray}
F_{x} = 2R \sin(\theta)  \quad
F_{y} = R \cos(\theta) \Rightarrow F_{x} = 2F_{y}\tan(\theta)
\end{eqnarray}

The force in the linear spring simply is $F_{y} = k_{l}f(x)$. The tangent of the profile $f(x)$ is the derivative at that point $\tan \theta = \dfrac{df(x)}{dx}$. With a profile of $f(x)=ax^2+b$ the Duffing characteristic follows:

\begin{equation}
F(x) = 2k_{l}f(x)\frac{df(x)}{dx} = 4k_{l}a(bx + ax^3) = kx + k_{3}x^3
\label{eq:spring_force}
\end{equation}

with $k = 4k_{l}ab$ and $k_{3}=4k_{l}a^2$.
The linear stiffness term is caused by a vertical shift ($b$) of the profile, pre-stressing the linear springs. Practically, it is difficult to precisely adjust $b$ and even a small shift in the linear term greatly improves the Duffing oscillator's performance as a vibration absorber (\cite{dekemele2018performance}), so identification of this parameter is critical.

\subsection{Experimental setup}

\begin{figure}[b]
	\centering
	\includegraphics[width=0.46\textwidth]{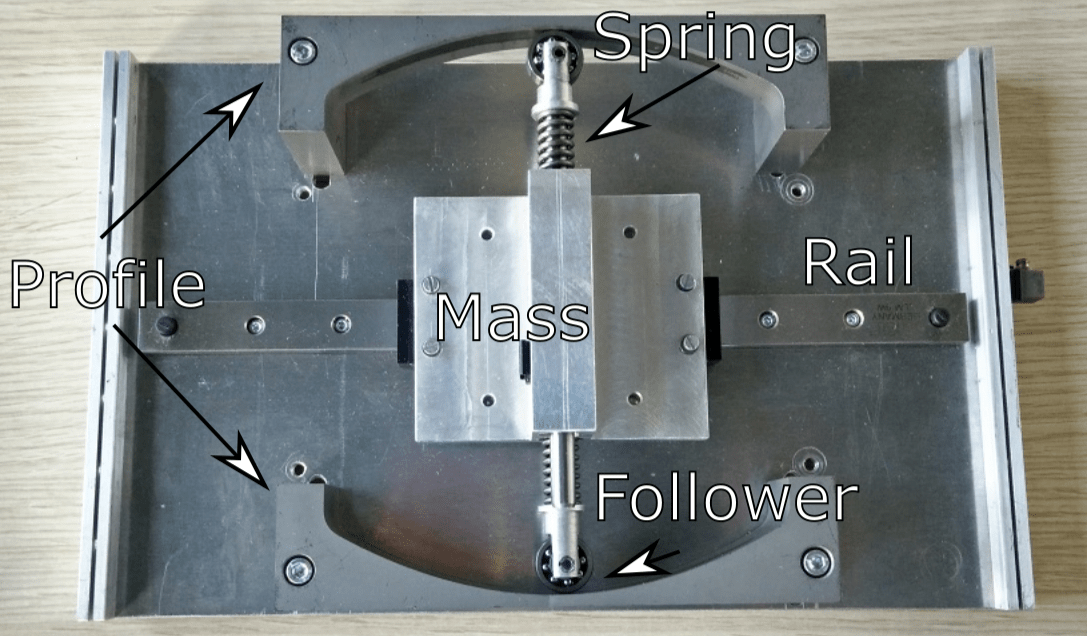}
	\caption{Realization of the Duffing oscillator design}
	\label{fig:realisation}
\end{figure}

The realization of the Duffing oscillator design is shown on Fig.~\ref{fig:realisation}. A mass of $0.49$ kg was fitted on a lubricated linear guide rail of $SKF$ with an unknown damping (also important for absorber performance), and two profiles were CNC-machined with $a=4$ $\textrm{m}^{-1}$. The rail and profiles were bolted to a plate. The followers are SKF roller bearing. The linear springs have a rated stiffness of $k_{l}=16.7$ kN which means the cubic stiffness is designed to be $k_{3}=1.07 $ $\textrm{MN/m}^3$.
The pre-stress, expressed as $b$ in equation~(\ref{eq:spring_force}) is unknown and should be revealed with the identification procedure. 

The Duffing oscillator is fitted on a shaking table which is a linear
permanent magnet motor from Beckhoff (AL2012) controlled by a Beckhoff PLC (CX5210). This shaking table will impose the ground excitation $z$ in equation~(\ref{eq:Duffing_ground}). Both the linear acceleration of the shaking table itself and the mass of the Duffing oscillator are measured with accelerometers. To acquire the velocity and displacement, the accelerations are integrated by using the proposed algorithm in \cite{Schoukens1990}, suitable given that the signals stay in a well-defined bandwidth. Here, the bandwidth is ensured by applying a sine sweep as ground displacement. 

\begin{figure}
	\centering
	\includegraphics[width=0.19\textwidth]{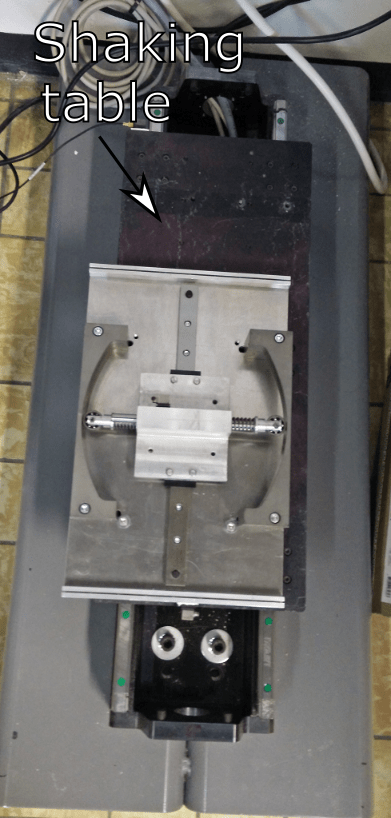}
	\caption{Duffing oscillator on shaking table}
	\label{fig:shaking_table}
\end{figure}

\section{Identification Algorithm}

Considering the relative ground coordinate $q(t)$ and its derivatives, $\dot{q}(t)$ and $\ddot{q}(t)$, the goal of the presented algorithm in this section is to reconstruct equation~(\ref{eq:Duffing_ground}) in the form of a state space model as in equation~(\ref{eq:tstatemodel}) with consideration of two states.
There are no primary assumptions made on the form of the model and both the structure and the values of the coefficients are derived from captured data.

\begin{gather} \label{eq:tstatemodel}
\left\{
\begin{matrix*}[l] \dot{q_{1}} = -\dfrac{c}{m} q_{1} -\dfrac{k}{m} q_{2} -\dfrac{k_{3}}{m} q_{2}^3 - \ddot{z} \\[10pt] \dot{q_{2}} = q_{1} \end{matrix*}
\right.
\end{gather}

In order to achieve this, a collection of preliminary functions of the states (equation~(\ref{eq:library})) is constructed using the measurement data. Each column represent a possible term for the right hand side of the model.
Concerning the establishment of this collection it is crucial to use the prior expert knowledge of the system to select the proper linear and nonlinear basis. 
It is common practice to use constants, polynomials and depending on the use case trigonometric functions as well. 
Additionally the time-varying inputs to the system are added to the collection. Therefore it is possible to recognize the effect of input on each state of the data.

\begin{multline} \label{eq:library}
\textbf{A(q,$\ddot{\textbf{z}}$)}
 = \\ \begin{pmatrix}
  | & | & | & | & | & | & | & | & | & | \\ 
\textbf{1} & \textbf{q} & \textbf{q}^{2} & \cdots & \sin(\textbf{q}) & \cos(\textbf{q}) & \cdots & \ddot{\textbf{z}} & \ddot{\textbf{z}}^2 & \cdots \\
  | & | & | & | & | & | & | & | & | & | \\
 \end{pmatrix}_{n\times m}
\end{multline}

The collection has n rows which equals to the number of time-varying data samples. The number of columns, m, depends on the selection of basis functions and the order of them, as well as the number of the states, p.

Discovery of the model architecture means that the correct terms from the collection should be chosen with the right coefficients. So far in the related literature the focus has been on finding the correct terms and the coefficients simultaneously, and the results of such an algorithm shown to be promising when applied on simulation data. 
The major drawback of this procedure is that selection of the terms is restricted because of finding the true coefficients and this limitation is more noticeable when identifying an experimental setup.
Consequently we present a modified two step algorithm for the sparse identification method, in which first the form of the equation with the right terms are selected and afterwards true coefficients for those terms are calculated.

For the sake of finding the model structure, we need to solve:

\begin{equation} \label{XI}
\dot{\textbf{q}} = \textbf{A(q,$\ddot{\textbf{z}}$)} \textbf{$\Xi$}
\end{equation}

to find the sparse matrix of active terms, \textbf{$\Xi$}, based on the assumption that only a few of the nonlinear terms actually appear in the governing equation. In other words, in the space of all possible nonlinear terms that can form the equation only a few of them have a considerable impact on the construction of the data and the model is a linear combination of these high-impact terms.

The distinct LASSO optimization problem from \cite{tibshirani1996regression} expressed by equation~(\ref{L1}) is solved to find the sparse vector of active terms for each column of \textbf{$\Xi$} matrix in equation~(\ref{XI}) for varying regularization parameter values $\lambda$.

\begin{align} \label{L1}
\textbf{$\Xi$}^{\ast} = \argminA_\Xi  \norm{\textbf{A} \textbf{$\Xi$} - \dot{\textbf{q}} }_{2} + \text{$ \lambda $}\norm{ \textbf{$\Xi$} }_{1}
\end{align}

The alternating direction method of multipliers algorithm (ADMM) is used to solve the LASSO problem in equation~(\ref{L1}). 
Each value of $\lambda$ results in a different model of the system. The degree of sparsity of the obtained model increases with the growth of $\lambda$. However as $\lambda$ rises, more terms are neglected and consequently the amount of error grows.
Hence, the selection of the correct sparse model is based on a trade-off between accuracy and complexity via Pareto front analysis. 
After electing the suitable model, we will proceed further by searching for the coefficients only for the terms that remain in the model.

The different steps of the procedure are explained in Algorithm~\ref{alg:ADMM}.
At each iteration (r), one model is being identified for a specific value of $\lambda$. The selection of the range this variation is arbitrary and defines the total number of iterations (s). The mean square error of each model is then calculated and later used in the Pareto front analysis.

\begin{algorithm}
\caption{ADMM LASSO Algorithm}
\label{alg:ADMM}
\begin{algorithmic}[1]
\Require Time-varying measurement data: {$\underset{n\times p}{\textbf{q}}$}, $\underset{n\times p}{\dot{\textbf{q}}}$, $\underset{n\times p}{\ddot{\textbf{q}}}$ and $\underset{n\times 1}{\ddot{\textbf{z}}}$
\Procedure{Two step ADMM-LASSO}{}
\State  $\underset{n\times m}{\textbf{A}(\textbf{q},\ddot{\textbf{z}})}$ \Comment{construct collection of candidate terms}
\For{r = 1 : s} \Comment{Changing the value of $\lambda$}
\State \textbf{Solve} \Comment{Find the r$_{th}$ model}
\begin{align*} \label{L1}
\textbf{$\Xi$}^{\ast}_{r} = \argminA_\Xi  \norm{\textbf{A} \textbf{$\Xi$} - \dot{\textbf{q}} }_{2} + \text{$ \lambda $}_{r} \norm{ \textbf{$\Xi$} }_{1}
\end{align*}  
\State \textbf{MSE} \Comment{Compute mean square error of the fit}
\begin{align*}
\text{MSE}_{r} = \frac{1}{n}\sum_{}^{} (\textbf{A}\textbf{$\Xi$}_{r}^{\ast}- \dot{\textbf{q}})^2
\end{align*} 
\EndFor
\State \textbf{Pareto front analysis} ($\textbf{$\Xi$}^{\ast}_{true}$) \Comment{Choose the sparsest model within the acceptable error range}
\For {i = 1 : m}
 \For {j = 1 :p}
  \If {$\underset{m\times p}{\textbf{$\Xi$}^{\ast}_{true}}$ $(i,j) \neq 0$}
   \State $\textbf{A} / \dot{\textbf{q}}$  \Comment{Compute the coefficients}
    \EndIf
 \EndFor
\EndFor
\EndProcedure
\end{algorithmic}
\end{algorithm}

\begin{multicols}{1}
\begin{strip}
\begin{equation} \label{eq:DuffingLibrary}
\textbf{A(q,$\ddot{\textbf{z}}$)}
 = \\ \begin{pmatrix}
  \rule{0pt}{3ex}
1 & \quad  q_{1}(t_{0})  & \quad q_{2}(t_{0}) & \quad q_{1}^2(t_{0}) & \quad q_{1}q_{2}(t_{0}) & \quad q_{2}^2(t_{0}) & \quad \cdots  &  \quad \ddot{z}(t_{0}) \\
\rule{0pt}{3ex}
1 & \quad q_{1}(t_{1})  & \quad q_{2}(t_{1}) & \quad q_{1}^2(t_{1}) & \quad q_{1}q_{2}(t_{1}) & \quad q_{2}^2(t_{1}) & \quad \cdots  & \quad \ddot{z}(t_{1})  \\
\rule{0pt}{3ex}
1 & \quad q_{1}(t_{2})  & \quad q_{2}(t_{2}) & \quad q_{1}^2(t_{2}) & \quad q_{1}q_{2}(t_{2}) & \quad q_{2}^2(t_{2}) & \quad \cdots  & \quad \ddot{z}(t_{2}) \\
\rule{0pt}{3ex}
 & \quad & \quad &  \quad\cdots &  \quad & \quad & \quad  \vdots  \\
 \rule{0pt}{3ex}
1 & \quad q_{1}(t_{n})  & \quad q_{2}(t_{n}) & \quad q_{1}^2(t_{n}) & \quad q_{1}q_{2}(t_{n}) & \quad q_{2}^2(t_{n}) & \quad \cdots  & \quad \ddot{z}(t_{n})\\
 \rule{0pt}{1.5ex}
 \end{pmatrix}
\end{equation}
\end{strip}
\end{multicols}

\section{Results and Discussion}

\begin{figure}
\begin{center}
\includegraphics[width=9.7cm]{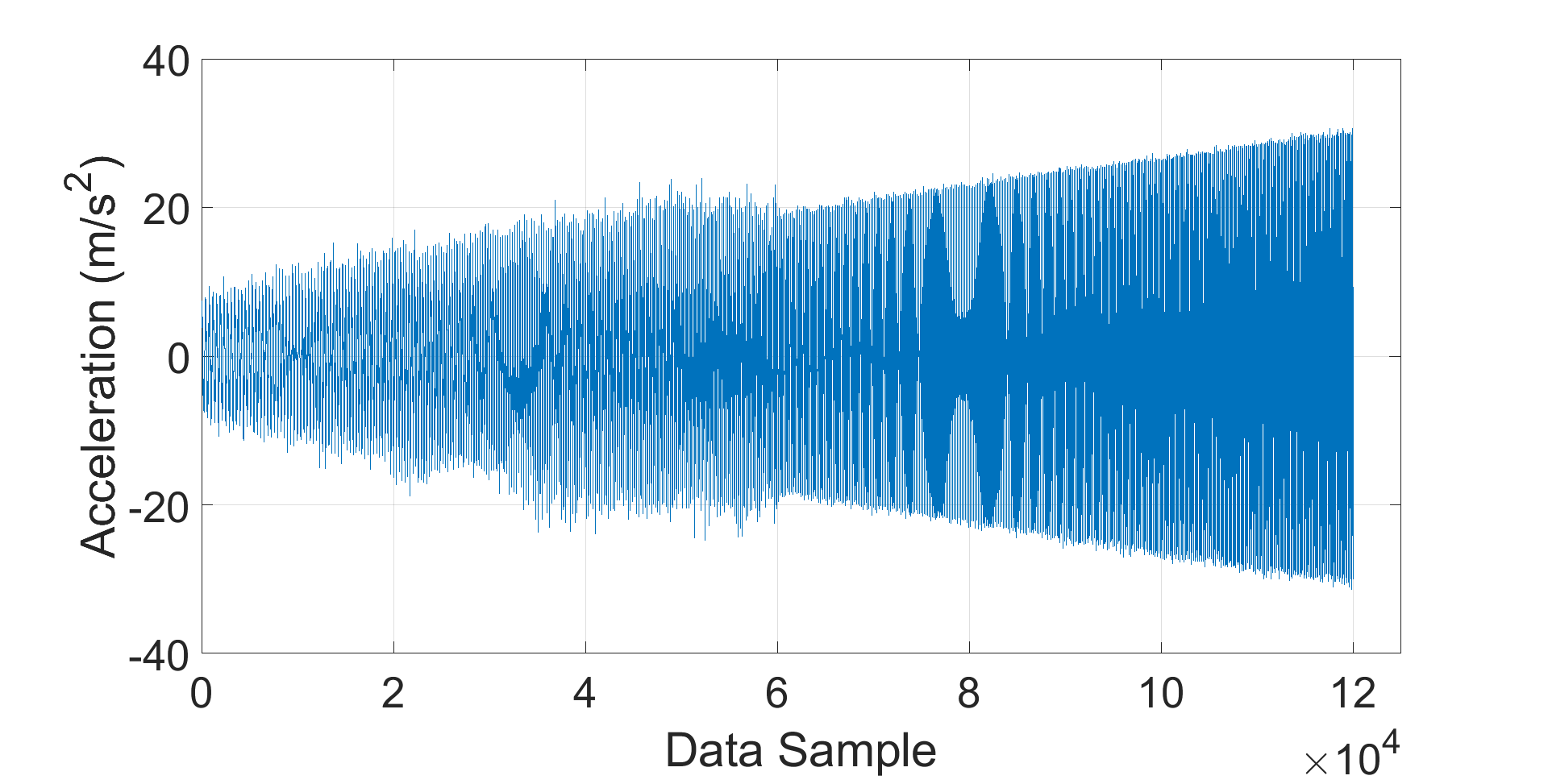}    
\caption{Input sweep} 
\label{fig:input}
\end{center}
\end{figure}

\begin{figure}
\begin{center}
\includegraphics[width=9.7cm]{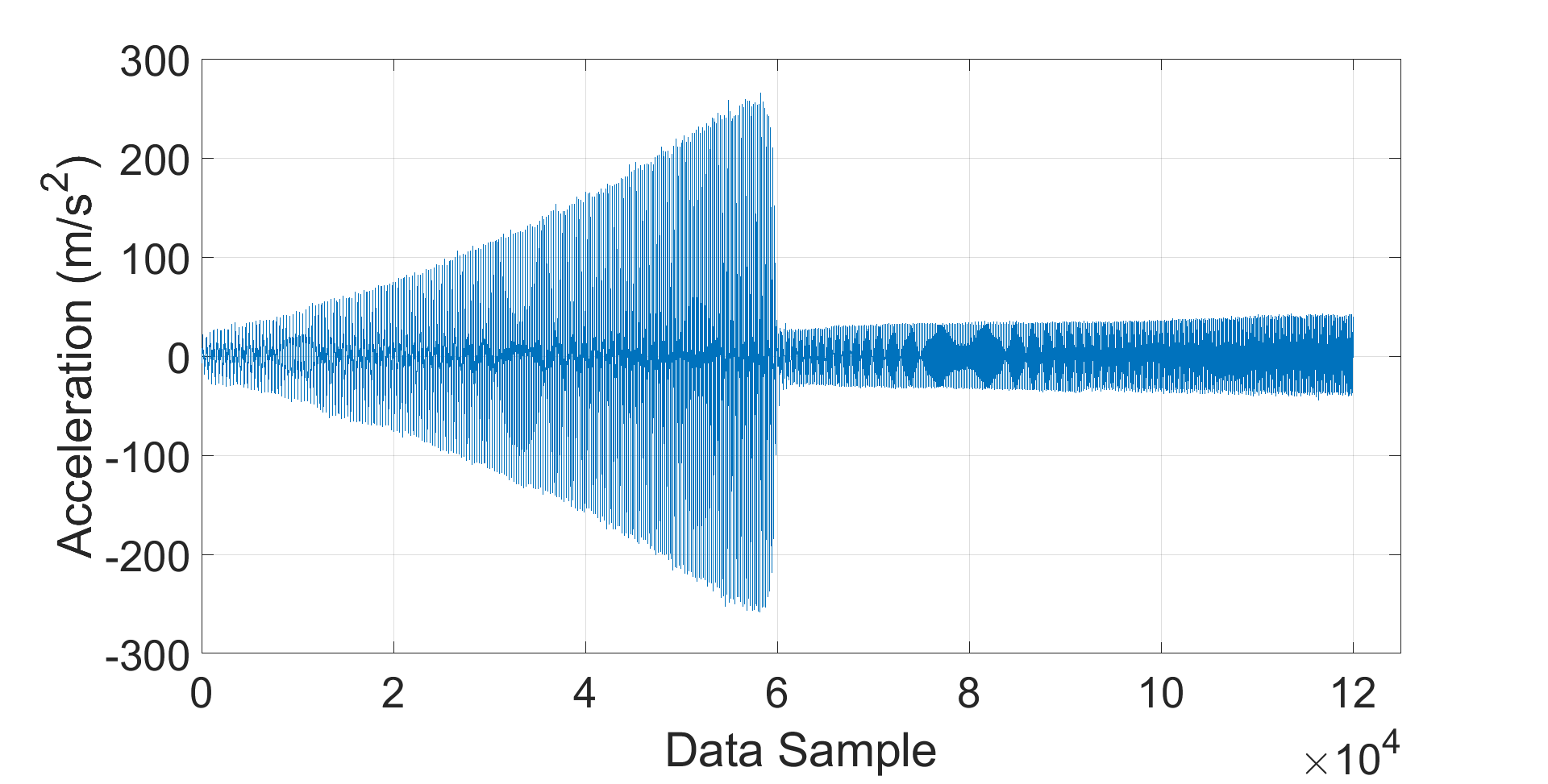}    
\caption{Output acceleration} 
\label{fig:out}
\end{center}
\end{figure}

In this section the results of applying the modified algorithm on the measurement data is presented and analyzed.
The input to the setup, namely the acceleration of the shaking table and the relative acceleration between the mass and shaking table as the output both with the sampling time $0.488$ ms can be seen in Fig.~\ref{fig:input} and Fig.~\ref{fig:out} respectively. The applied input is a linearly varying frequency sweep between 1 and 20 Hz. 
 The first $30000$ data samples are selected for validation and the remainder are used for identification. 

As expressed in equation~(\ref{eq:DuffingLibrary}), in this paper we use constants, polynomials to the fourth order and the input of the system to build the collection of possible terms for construction of the model.
Moreover, the selected total number of iterations (s) is 100.

In order to select the correct model of the system the balance between sparsity represented by the number of terms in the equation and the accuracy represented by mean square error of the fit should be found. Hence we limit the error tolerance to $+25\%$ of the minimum mean squared error (MSE) and select the sparsest solution in this span.

\begin{figure}
\begin{center}
\includegraphics[width=9.7cm]{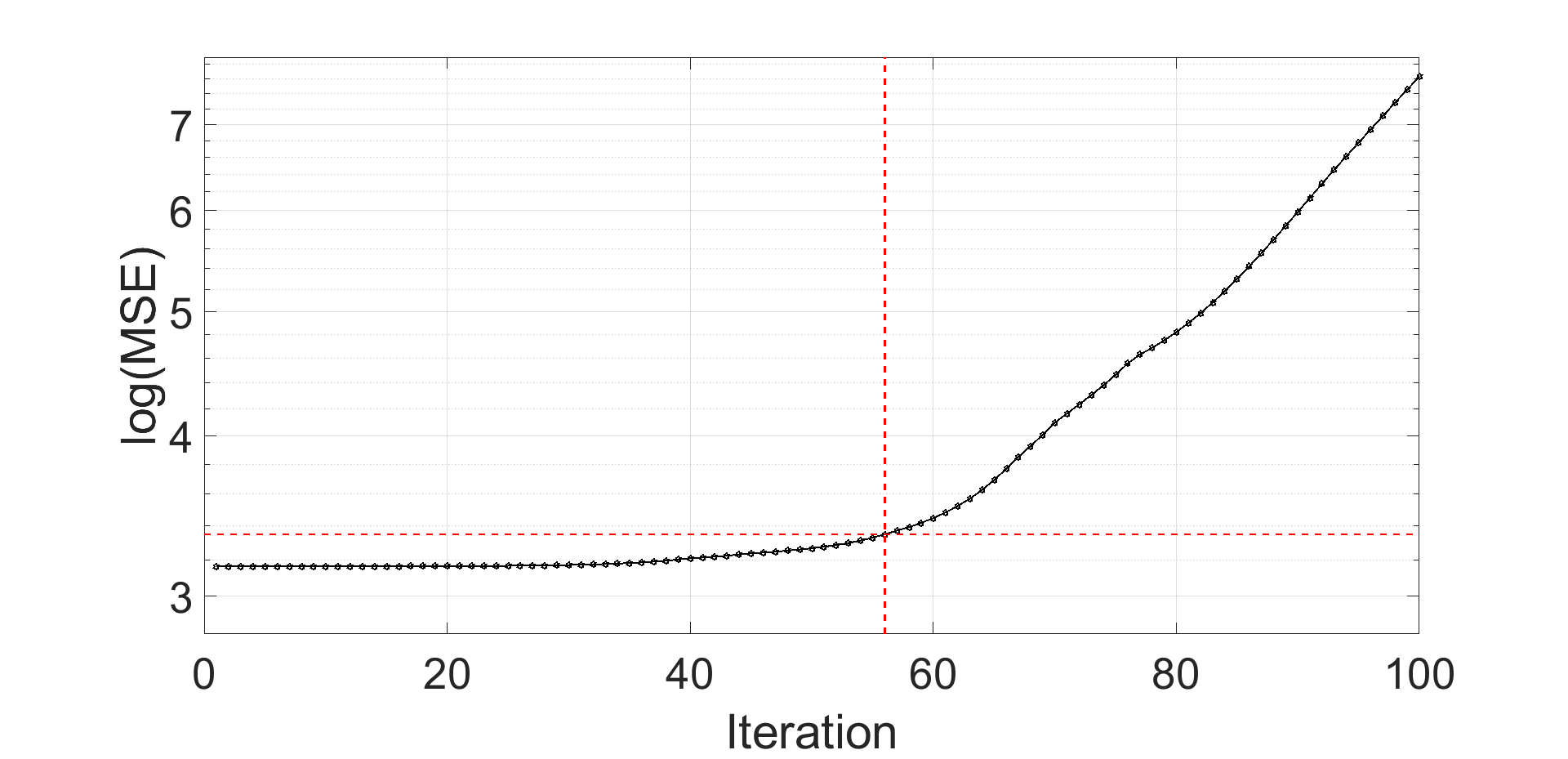}    
\caption{Log mean square error (MSE) at each iteration} 
\label{fig:Pareto}
\end{center}
\end{figure}

Fig.~\ref{fig:Pareto} illustrates such an analysis in case of $\dot{q}_{1}$. 
The horizontal axis shows the number of iteration (r) which reports the increase of regularization parameter $\lambda$ that results in a more sparse solution by eliminating less important terms. The horizontal red line illustrates the error tolerance limit. It can be seen that in case of the Duffing setup, the 56$_{th}$ iteration gives the sparsest model in the acceptable error range.

\begin{figure*}
\begin{center}
\includegraphics[width=17.2cm]{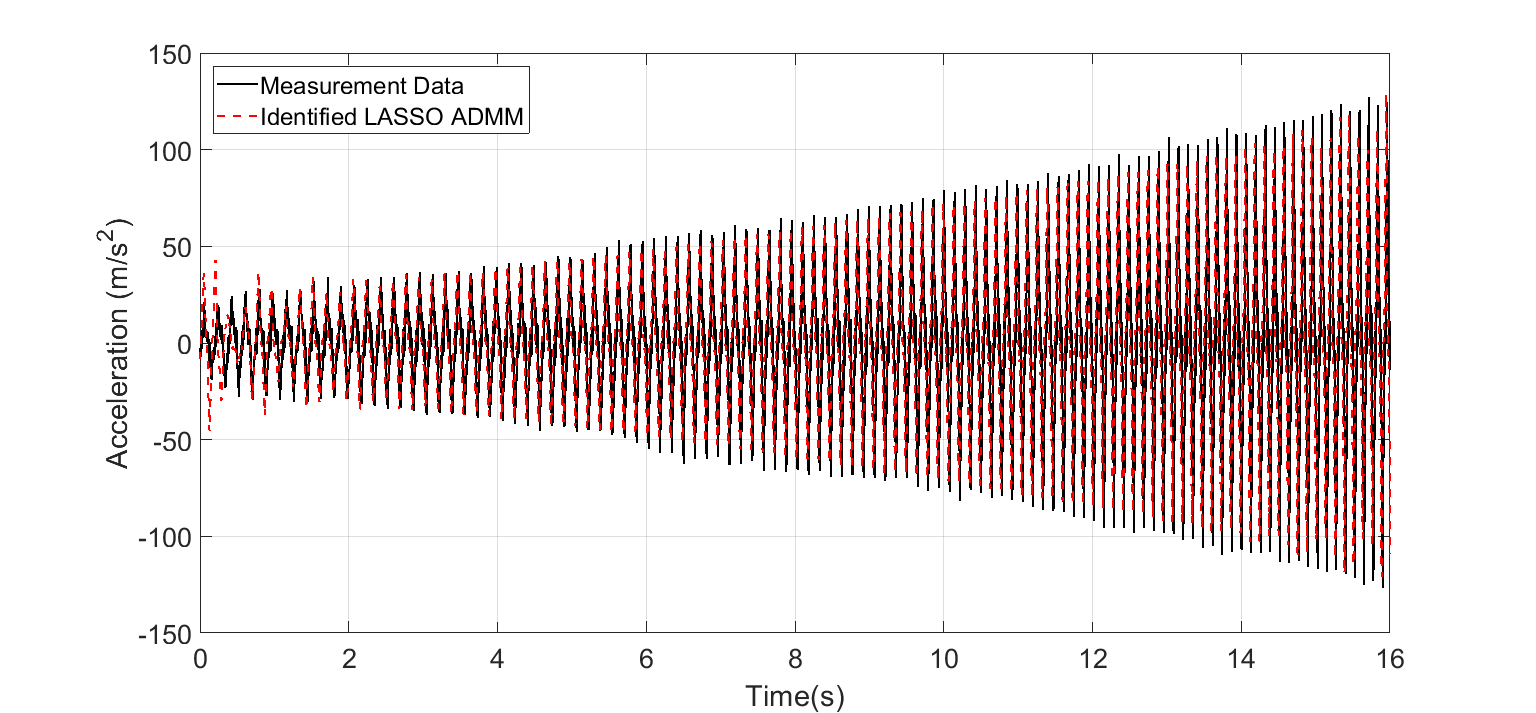}    
\caption{Validation of the identified system with the measurement data} 
\label{fig:compareacc3}
\end{center}
\end{figure*}

Table~\ref{tb:ADMMcoef} presents the discovered dynamic model of the system whereas Table~\ref{tb:Bruncoef} shows the equation derived from the sequential thresholded least-squares algorithm by \cite{brunton2016sparse} ($\lambda$ = 0.95).
By comparing these two models, the advantages of the proposed algorithm are clear.

Regarding the model in Table~\ref{tb:ADMMcoef} it can be observed that along with the terms that we have anticipated based on equation~(\ref{eq:Duffing_ground}), an additional nonlinear term is present as $q_{1}q_{2}q_{2}$.
This term is a damping term depending on the displacement. It appears as the force on the follower bearings depend on the position, namely the reaction force of the profile on the bearing. Additionally, a study of other nonpolynomial nonlinear damping effects should be investigated to fully capture the damping behavior typically associated with realizations of mechanical devices.

\begin{table}[t]
\begin{center}
\caption{Identified model via two step alternating direction method of multipliers algorithm}
\label{tb:ADMMcoef}
\begin{tabular}{lccc}
 & $\dot{q_{1}}$ & $\dot{q_{2}}$ \\\hline
  \rule{0pt}{2.5ex}
$1$ & 0 & 0 \\
 \rule{0pt}{2.5ex}
$q_{1}$ & -3.7294 & 1.0000 \\ 
 \rule{0pt}{2.5ex}
$q_{2}$ & -971.0169 & 0 \\
 \rule{0pt}{2.5ex}
$q_{1}q_{1}$ & 0 & 0 \\
 \rule{0pt}{2.5ex}
$q_{1}q_{2}$ & 0 & 0 \\
 \rule{0pt}{2.5ex}
$q_{2}q_{2}$ & 0 & 0 \\
 \rule{0pt}{2.5ex}
$q_{1}q_{1}q_{1}$ & 0 & 0 \\
 \rule{0pt}{2.5ex}
$q_{1}q_{1}q_{2}$ & 0 & 0 \\
 \rule{0pt}{2.5ex}
$q_{1}q_{2}q_{2}$ & 1.1620e+04 & 0 \\
 \rule{0pt}{2.5ex}
$q_{2}q_{2}q_{2}$ & -2.5033e+06 & 0 \\
 \rule{0pt}{2.5ex}
$q_{1}q_{1}q_{1}q_{1}$ & 0 & 0 \\
 \rule{0pt}{2.5ex}
$q_{1}q_{1}q_{1}q_{2}$ & 0 & 0 \\
 \rule{0pt}{2.5ex}
$q_{1}q_{1}q_{2}q_{2}$ & 0 & 0 \\
 \rule{0pt}{2.5ex}
$q_{1}q_{2}q_{2}q_{2}$ & 0 & 0 \\
 \rule{0pt}{2.5ex}
$q_{2}q_{2}q_{2}q_{2}$ & 0 & 0 \\
 \rule{0pt}{2.5ex}
$\ddot{z}$ &  -1.0243 & 0 \\  \hline
\end{tabular}
\end{center}
\end{table}

\begin{table}[t]
\begin{center}
\caption{Identified model via sequential thresholded least-squares algorithm}
\label{tb:Bruncoef}
\begin{tabular}{lccc}
 & $\dot{q_{1}}$ & $\dot{q_{2}}$ \\\hline
 \rule{0pt}{2.5ex}
$1$ & 0 & 0 \\
 \rule{0pt}{2.5ex}
$q_{1}$ & -3.7055 & 1.0000 \\ 
 \rule{0pt}{2.5ex}
$q_{2}$ & -995.0195 & 0 \\
 \rule{0pt}{2.5ex}
$q_{1}q_{1}$ & 0 & 0 \\
 \rule{0pt}{2.5ex}
$q_{1}q_{2}$ & 50.6891 & 0 \\
 \rule{0pt}{2.5ex}
$q_{2}q_{2}$ & -202.2274 & 0 \\
 \rule{0pt}{2.5ex}
$q_{1}q_{1}q_{1}$ & 0 & 0 \\
 \rule{0pt}{2.5ex}
$q_{1}q_{1}q_{2}$ & 15.2592 & 0 \\
 \rule{0pt}{2.5ex}
$q_{1}q_{2}q_{2}$ & 1.1547e+04 & 0 \\
 \rule{0pt}{2.5ex}
$q_{2}q_{2}q_{2}$ & -2.4946e+06 & 0 \\
 \rule{0pt}{2.5ex}
$q_{1}q_{1}q_{1}q_{1}$ & 0 & 0 \\
 \rule{0pt}{2.5ex}
$q_{1}q_{1}q_{1}q_{2}$ & -2.7156 & 0 \\
 \rule{0pt}{2.5ex}
$q_{1}q_{1}q_{2}q_{2}$ & -846.6316 & 0 \\
 \rule{0pt}{2.5ex}
$q_{1}q_{2}q_{2}q_{2}$ & -1.5241e+03 & 0 \\
 \rule{0pt}{2.5ex}
$q_{2}q_{2}q_{2}q_{2}$ & 5.7676e+04 & 0 \\
 \rule{0pt}{2.5ex}
$\ddot{z}$ &  -1.0208 & 0 \\  \hline\end{tabular}
\end{center}
\end{table}

Moreover, there are several sources of error including the need for numerical differentiation and integration which may introduce errors in the estimation of signals as well as uncertainties of the accelerometers.

Nevertheless as it is apparent from Fig.~\ref{fig:compareacc3} the correlation between the results and the measurement data validates the model. Additionally, the estimated cubic coefficient, the term $q_{2}q_{2}q_{2}$ or k$_{3}$/m is estimated as $2.5$ MN/m$^3$/kg, close to the designed coefficient of $2.18$ MN/m$^3$/kg.

\section{Conclusion}

In this paper we have identified a laboratory designed mechanical Duffing setup for passive vibration control through sparse regression procedure. Our work has led us to conclude that in the case of measurement data some adjustments to the current sparse identification algorithms are crucial. The used alternating direction method of multipliers has shown promising results with respect to the system identification of the nonlinear Duffing oscillator starting from actual measurement data. Accordingly we devised a strategy which ensures achieving the sparse state space model of the system in addition to finding hidden nonlinear terms such as friction, often present in real mechanical systems. These findings add to a growing body of literature on nonlinear dynamic system identification and suggests consideration of measurement data for testing developed algorithms as a vital issue for future research.

\begin{ack}
This work was supported by the strategic basic research project EMODO and the ICON project Multi-Sensor of Flanders Make, the Strategic Research Centre for the Manufacturing Industry; and the FWO research project G.0D93.16N.

We would like to thank Michiel Dhont for his assistants in capturing and processing the measurements.
\end{ack}

\bibliography{CHAOS2018}             
                                                  
\end{document}